\documentclass[10pt,conference]{IEEEtran}

\usepackage[dvips]{graphicx}
\usepackage{psfrag}
\usepackage[notrantools]{research7}
\allowdisplaybreaks

\newcounter{mytempequationscounter}

\author{%
  \authorblockN{Amos Lapidoth}
  \authorblockA{
    ETH Zurich\\
    Zurich, Switzerland \\
    Email: lapidoth@isi.ee.ethz.ch}  
  \and
  \authorblockN{Stefan M.~Moser}
  \authorblockA{
    National Chiao Tung University (NCTU)\\
    Hsinchu, Taiwan \\
    Email: stefan.moser@ieee.org}
  \and
  \authorblockN{Mich\`ele A.~Wigger}
  \authorblockA{
   ETH Zurich\\
    Zurich, Switzerland \\
    Email: wigger@isi.ee.ethz.ch}  
}

\title{On the Capacity of Free-Space Optical
  Intensity Channels}

\begin{document}

\maketitle

\begin{abstract}
  New upper and lower bounds are presented on the capacity of the
  free-space optical intensity channel. This channel is characterized
  by inputs that are nonnegative (representing the transmitted optical
  intensity) and by outputs that are corrupted by additive white
  Gaussian noise (because in free space the disturbances arise from
  many independent sources).  Due to battery and safety reasons the
  inputs are simultaneously constrained in both their average and peak
  power. For a fixed ratio of the average power to the peak power the
  difference between the upper and the lower bounds tends to zero as
  the average power tends to infinity, and the ratio of the upper and
  lower bounds tends to one as the average power tends to zero.
  
  The case where only an average-power constraint is imposed on the
  input is treated separately. In this case, the difference of the
  upper and lower bound tends to 0 as the average power tends to
  infinity, and their ratio tends to a constant as the power tends
  to zero.
\end{abstract}

\section{Introduction}
\label{sec:3intro}

We consider a channel model for short
optical communications in free space such as the communication between a
remote control and a TV. We assume a channel model based on
\emph{intensity modulation} where the signal is modulated onto the
optical intensity of the emitted light.  Thus, the channel input is
proportional to the light intensity and is therefore nonnegative.  We
further assume that the receiver directly measures the incident
optical intensity of the incoming signal, \emph{i.e.}, it produces an
electrical current at the output which is proportional to the detected
intensity. Since in ambient light conditions the received signal is
disturbed by a high number of independent sources, we model the noise
as Gaussian.  Moreover, we assume that the
line-of-sight component is dominant and ignore any effects due to
multiple-path propagation like fading or inter-symbol interference.

Optical communication is restricted not only by battery power, but
also for safety reasons by the maximum allowed peak power. We
therefore assume simultaneously two constraints: an average-power
constraint $\Es$ and a maximum allowed peak power $\amp$. The
situation where only a peak-power constraint is imposed corresponds to
$\Es = \amp$. The case of only an average-power constraint is treated
separately.

In this work we study the channel capacity of such an optical
communication channel and present new upper and lower bounds. The
maximum gap between upper and lower bound never exceeds 1 nat when the
ratio of the average-power constraint to the peak-power constraint is
larger than $0.03$ or when only the average power is
constrained but not the peak power.  Asymptotically when the available
average and peak power tend to infinity with their ratio held fixed,
the upper and lower bounds coincide, \emph{i.e.}, their difference
tends to 0. We also present the asymptotic behavior of
the channel capacity  in the limit when the power tends to 0. 

The channel model has been studied before, \emph{e.g.}, in
\cite{chanhranilovickschischang05_1} and is described in detail in the
following. The received signal is corrupted by additive noise due to
strong ambient light that causes high-intensity shot noise in the
electrical output signal. In a first approximation this shot noise can
be assumed to be independent of the signal itself, and since the noise
is caused by many independent sources it is reasonable to model it as
an independent and identically distributed (IID) Gaussian process.
Also, without loss of generality we assume the noise to be zero-mean,
since the receiver can always subtract or add any constant signal.

Hence, the channel output ${Y}_k$ at time $k$, modeling a sample of
the electrical output signal, is given by
\begin{equation}\label{eq:3law}
  {Y}_k = x_k + {Z}_k,
\end{equation}
where $x_k\in \Reals_0^{+}$ denotes the time-$k$ channel input and
represents a sample of the electrical input current that is
proportional to the optical intensity and therefore nonnegative, and
where the random process $\{Z_k\}$ modeling the additive noise is
given by
\begin{equation}\label{eq:3noise}
  \{{Z}_k\} \sim \text{IID } \NormalR{0}{\sigma^2}.
\end{equation}
It is important to note that, unlike the input $x_k$, the output $Y_k$
may be negative since the noise introduced at the receiver can be
negative.



Since the optical intensity is proportional to the optical power, in
such a system the instantaneous optical power is proportional to the
electrical input current \cite{kahnbarry97_1}. This is in contrast to
radio communication where usually the power is proportional to
the square of the input current. Therefore, in addition to the
implicit nonnegativity constraint on the input,
\begin{equation}
\label{eq:non-neg}
X_k \geq 0,
\end{equation} we
assume constraints both on the peak and the average power, \emph{i.e.}, 
\begin{IEEEeqnarray}{c}
  \Prv{ X_k > \amp} =0,   \label{eq:3defpp}\\
  \E{X_k} \le \Es.  \label{eq:3defavg} 
\end{IEEEeqnarray}

We shall denote the ratio
between the average power and the peak power by $\alpha$,
\begin{equation}
  \label{eq:3defalpha}
  \alpha \eqdef \frac{\Es}{\amp},
\end{equation}
where we assume $0 < \alpha \le 1$. Note that $\alpha=1$
corresponds to the case with only a peak-power constraint. Similarly,
$\alpha \ll 1$ corresponds to a dominant average-power constraint and
only a very weak peak-power constraint.

We denote the capacity of the described channel with peak-power
constraint $\amp$ and average-power constraint $\Es$  by
$\const{C}(\amp,\Es)$. The capacity is given by \cite{shannon48_1}
\begin{equation}\label{eq:C}
\const{C}(\amp,\Es)= \sup I(X;Y)
\end{equation}
where $I(X;Y)$ stands for the mutual information
between the channel input $X$ and the 
channel output $Y$, where conditional on the 
input $x$ the output $Y$ is Gaussian $\sim\Normal{x}{\sigma^2}$;
and where the supremum is over all laws on $X\geq 0$ satisfying $\Prv{X
\geq \amp}=0$ and $\E{X}\leq \Es$.

In the case of only an average-power constraint the capacity is
denoted by $\const{C}(\Es)$. It is given as in \eqref{eq:C}
except that the supremum is taken over all laws on $X\geq0$ satisfying
$\E{X}\leq \Es$.

The derivation of the upper bounds is based on a technique introduced
in \cite{lapidothmoser03_3}.  There a dual expression of mutual
information is used to show that for any channel law $W(\cdot|\cdot)$
and for an arbitrary distribution $R(\cdot)$ over the channel output
alphabet, the channel capacity is upper-bounded by
\begin{equation}
  \const{C} \leq \E[Q^*]{D \bigl( W(\cdot|X) \big\| R(\cdot) \bigr)}.
  \label{eq:upperbound1} 
\end{equation}
Here, $D(\cdot \| \cdot)$ stands for the relative entropy
\cite[Ch.~2]{coverthomas06_1}, and $Q^*(\cdot)$ denotes the
capacity-achieving input distribution. For more details about this
technique and for a proof of \eqref{eq:upperbound1}, see
\cite[Sec.~V]{lapidothmoser03_3}, \cite[Ch.~2]{moser04_1}. The
challenge of using \eqref{eq:upperbound1} lies in a clever choice of
the arbitrary law $R(\cdot)$ that will lead to a good upper bound.
Moreover, note that the bound \eqref{eq:upperbound1} still contains an
expectation over the (unknown) capacity-achieving input distribution
$Q^*(\cdot)$. To handle this expectation we will need to resort to some
further bounding like, \emph{e.g.}, Jensen's inequality
\cite[Ch.~2.6]{coverthomas06_1}.

The derivation of the firm lower bounds relies on the entropy power
inequality \cite[Th.~17.7.3]{coverthomas06_1}. 

The asymptotic results at high power follow directly by evaluation of
the firm upper and lower bounds. For the low power regime we introduce
an additional lower bound which does not rely on the entropy power
inequality. This lower bound is obtained by choosing a binary input
distribution, a choice which was inspired by \cite{smith71}, and by
then evaluating the corresponding mutual information.  For the cases
involving a peak-power constraint we further resort to the results on
the asymptotic expression of mutual information for weak signals in
\cite{perlovvandermeulen93_1}.

The results of this paper are partially based on the results in
\cite{wigger03_1} and \cite[Ch.~3]{moser04_1}.

The remainder is structured as follows. In the subsequent section we
state our results 
and  in Section~\ref{sec:derivation} we give a brief outline
of some of the derivations.



\section{Results}
\label{sec:3resultsone}

We start with an auxiliary lemma which is based on the symmetry of the
channel law and the concavity of channel capacity in the input
distribution.
\begin{lemma}
  \label{lem:inactiveaveragepower}
  Consider a peak-power constraint $\amp$ and an average-power
  constraint $\Es$ such that $\alpha =\frac{\amp}{\Es} > \frac{1}{2}$.
  Then the optimal input distribution\footnote{It was shown in
    \cite{chanhranilovickschischang05_1} that the optimal input
    distribution in \eqref{eq:C} is unique.}  in \eqref{eq:C}
  has an average power equal to half the peak power
  \begin{equation}
    \frac{\E[Q^*]{X}}{\amp} = \frac{1}{2},
  \end{equation}
  irrespective of $\alpha$. \emph{I.e.}, the average-power constraint
  is inactive for all $\alpha\in\left(\frac{1}{2}, 1\right]$, and in
  particular
  \begin{equation}
    \const{C}(\amp,\alpha \amp)=\const{C}\left(\amp,
      \frac{\amp}{2}\right),\quad \frac{1}{2}<\alpha\leq 1. 
  \end{equation}
\end{lemma}

We are now ready to state our results. We will distinguish between
three different cases: in the first two cases we impose on the input
both an average- and a peak-power constraint: in the first case the
average-to-peak power ratio $\alpha$ is restricted to lie in
$\left(0,\frac{1}{2}\right)$, and in the second case it is restricted
to lie in $\left[\frac{1}{2}, 1\right]$. (Note that by
Lemma~\ref{lem:inactiveaveragepower}, $\frac{1}{2}<\alpha\leq 1$
represents the situation with an inactive average-power constraint.)
In the third case we impose on the input an average power constraint
only.

In all three cases we present firm upper and lower bounds on the
channel capacity. The difference of the upper and lower bounds tends
to 0 when the available average and peak power tend to infinity with
their ratio held constant at $\alpha$. Thus we can derive the
asymptotic capacity at high power exactly. We also present the
asymptotics of capacity at low power: for the cases involving a
peak-power constraint we are able to state them exactly, and for the
case of only an average-power constraint we give the asymptotics up to
a constant factor.

\subsection{Bounds on Channel Capacity with both an Average- and a
  Peak-Power Constraint ($0 < \alpha < \frac{1}{2}$)}
\label{sec:strongresults}

\begin{theorem}
  \label{thm:strongresults}
  If $0<\alpha < \frac{1}{2}$, then $\const{C}(\amp,\alpha \amp)$ is
  lower-bounded by
  \begin{equation}
    \const{C}(\amp,\alpha\amp) \ge \frac{1}{2} \log \left(
      1+\amp^2\frac{e^{2 \alpha \mu^{\ast}}}{2 \pi e \sigma^2}  
      \left(\frac{1-e^{-\mu^{\ast}}}{\mu^{\ast}} \right)^2 \right),
    \label{eq:case1lower}
  \end{equation}
  and upper-bounded by each of the two bounds
  \begin{IEEEeqnarray}{rCl}
    \const{C}(\amp,\alpha\amp) & \le &
    \frac{1}{2}\log\left(1+\alpha(1-\alpha)\frac{\amp^2}{\sigma^2}\right),
    \label{eq:case1upperlow} 
    \\
    \const{C}(\amp,\alpha\amp) & \le &
    \left(1-\Qf{\frac{\delta+\alpha\amp}{\sigma}}-
      \Qf{\frac{\delta+(1-\alpha)\amp}{\sigma}} \right) \cdot
    \nonumber \\
    && \qquad \cdot \log \left(
      \frac{\amp}{\sigma}\cdot \frac{e^{ \frac{\mu \delta}{
            \amp}} -
        e^{-\mu\left(1+\frac{\delta}{\amp}\right)}}
      {\sqrt{2\pi}\mu\left(1-2\Qf{\frac{\delta}{\sigma}}\right)}
    \right)
    \nonumber \\
    && -\:\frac{1}{2} + \Qf{\frac{\delta}{\sigma}} +
    \frac{\delta}{\sqrt{2\pi}\sigma} e^{-\frac{\delta^2}{2\sigma^2}}
    \nonumber \\
    && +\:
    \frac{\sigma}{\amp}\frac{\mu}{\sqrt{2\pi}}
    \left(e^{-\frac{\delta^2}{2\sigma^2}}
      -e^{-\frac{\left(\amp+\delta\right)^2}{2\sigma^2}}\right)
    \nonumber \\
    && +\:
    \mu\alpha
    \left(1-2\Qf{\frac{\delta+\frac{\amp}{2}}{\sigma}}\right).
    \label{eq:case1upperhigh} 
  \end{IEEEeqnarray}
  Here $\Qf{\cdot}$ denotes the $\Q$-function defined by
  \begin{equation}
    \label{eq:defqfunction}
    \Qf{\xi} \eqdef \int_{\xi}^{\infty} \frac{1}{\sqrt{2\pi}}
    \cdot e^{-\frac{t^2}{2}} \d t, \qquad \forall \:\xi\in\Reals;
  \end{equation}
  $\mu > 0$ and $\delta > 0$ are free parameters; and 
  $\mu^{\ast}$ is the unique solution of
  \begin{equation}
    \label{eq:3defmugurk}
    \alpha = \frac{1}{\mu^{\ast}} -
    \frac{e^{-\mu^{\ast}}}{1-e^{-\mu^{\ast}}}.
  \end{equation}
\end{theorem}

\begin{figure}[tb]
  \centering
  \psfrag{A [dB]}[tc][bc]{$\frac{\amp}{\sigma}$ [dB]}
  \psfrag{C [nats]}[bc][tc]{$\const{C}$ [nats per channel use]}
  \psfrag{Strong Average- and Peak-Power Constraints, alpha=0.1}[cb][cb]{}
  \psfrag{lower bound                                                       .}[lc][lc]{\tiny lower bound \eqref{eq:case1lower}, $\alpha{=}0.1$}
  \psfrag{first upper bound}[lc][lc]{\tiny upper bound \eqref{eq:case1upperlow}, $\alpha{=}0.1$}
  \psfrag{second upper bound}[lc][lc]{\tiny upper bd.~\eqref{eq:case1upperhigh}, $\alpha{=}0.1$, using \eqref{eq:case1choicedelta}, \eqref{eq:case1choicemu}}
  \includegraphics[width=0.44\textwidth]{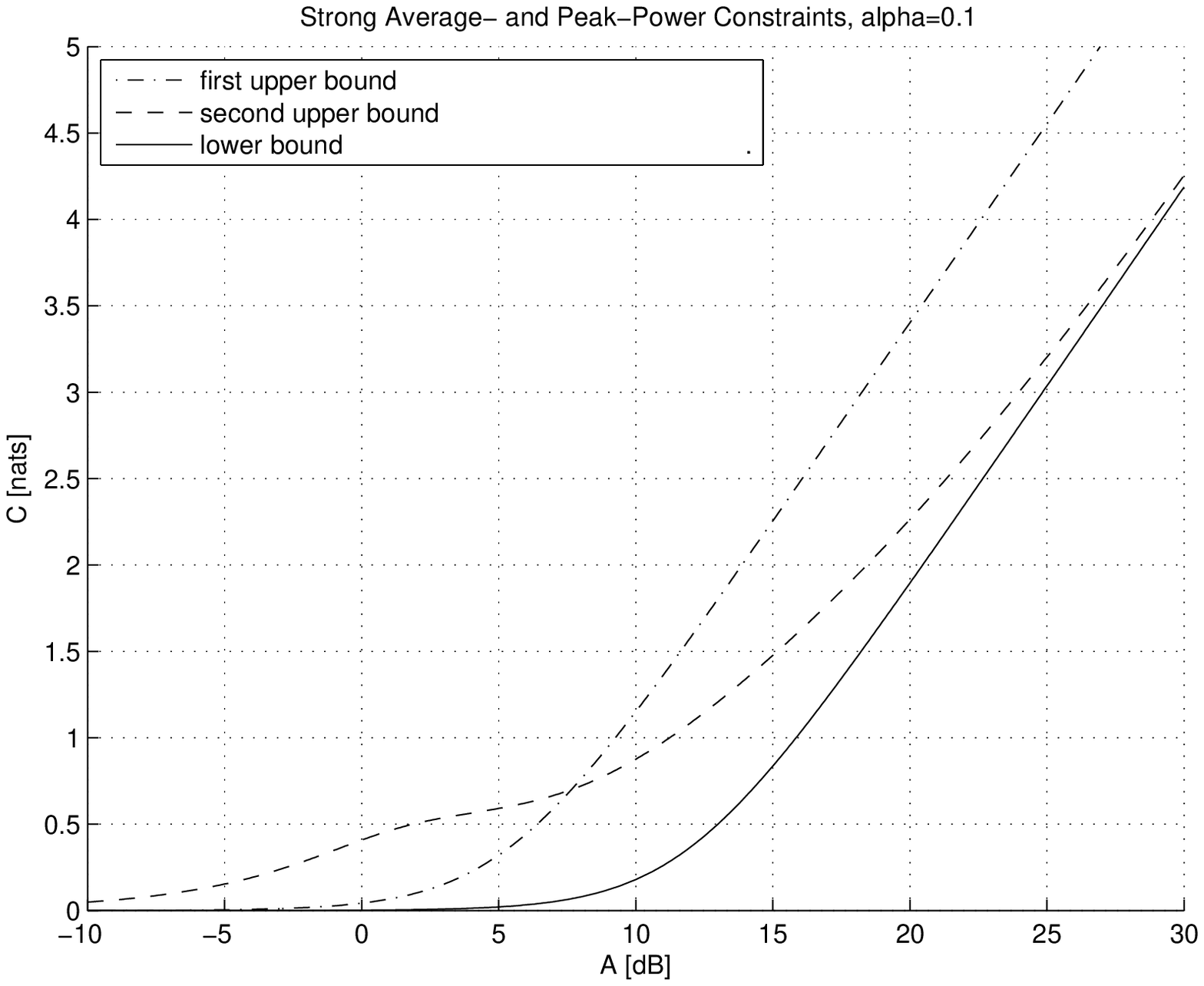}
  \caption{Bounds of Theorem~\ref{thm:strongresults} for a choice of
    the average-to-peak power ratio $\alpha=0.1$. The free parameters
    have been chosen as suggested in \eqref{eq:case1choicedelta} and
    \eqref{eq:case1choicemu}. The maximum gap between upper and lower
    bound is 0.72 nats (for $\amp/\sigma\approx 11.8 \text{ dB}$).}
  \label{fig:strongbounds2} 
\end{figure}

\begin{figure}[tb]
  \centering
  \psfrag{A [dB]}[tc][bc]{$\frac{\amp}{\sigma}$ [dB]}
  \psfrag{C [nats]}[bc][tc]{$\const{C}$ [nats per channel use]}
  \psfrag{Strong Average- and Peak-Power Constraints, alpha=0.4}[cb][cb]{}
  \psfrag{lower bound                                                       .}[lc][lc]{\tiny lower bound \eqref{eq:case1lower}, $\alpha{=}0.4$}
  \psfrag{first upper bound}[lc][lc]{\tiny upper bound \eqref{eq:case1upperlow}, $\alpha{=}0.4$}
  \psfrag{second upper bound}[lc][lc]{\tiny upper bd.~\eqref{eq:case1upperhigh}, $\alpha{=}0.4$, using \eqref{eq:case1choicedelta}, \eqref{eq:case1choicemu}}
  \includegraphics[width=0.44\textwidth]{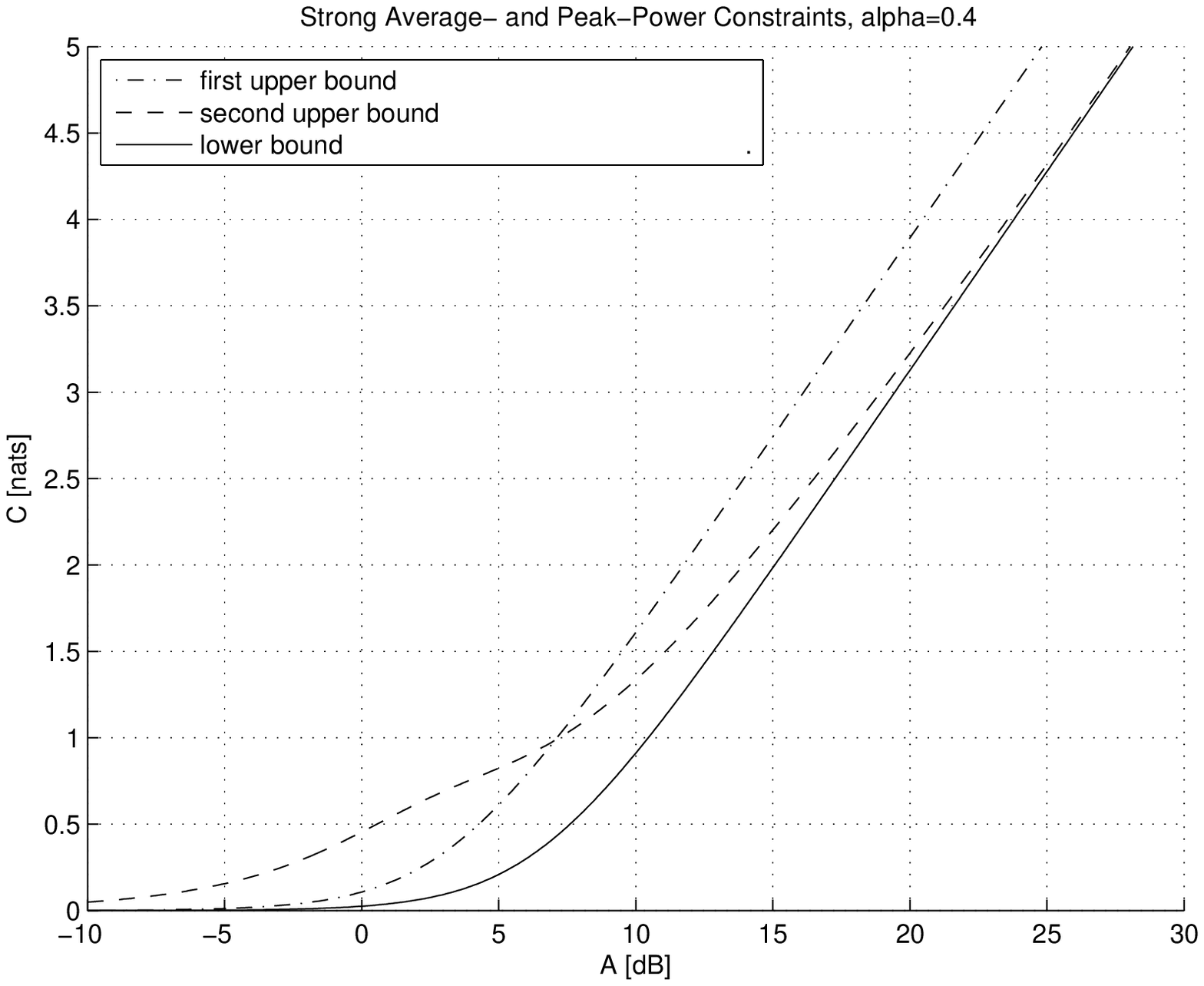}
  \caption{Bounds of Theorem~\ref{thm:strongresults} for a choice of
    the average-to-peak power ratio $\alpha=0.4$. The free parameters
    have been chosen as suggested in \eqref{eq:case1choicedelta} and
    \eqref{eq:case1choicemu}. The maximum gap between upper and lower
    bound is 0.56 nats (for $\amp/\sigma\approx 7.1 \text{ dB}$).}
  \label{fig:strongbounds3} 
\end{figure}

Note that $\mu^*$ is well-defined as the function $\mu^* \mapsto
\frac{1}{\mu^{\ast}} - \frac{e^{-\mu^{\ast}}}{1-e^{-\mu^{\ast}}}$ is
strictly monotonically decreasing over $(0,\infty)$ and tends to
$\frac{1}{2}$ for $\mu^*\downarrow 0$ and to $0$ for $\mu^* \uparrow
\infty$.

A suboptimal but useful choice for the free parameters
in the upper bound \eqref{eq:case1upperhigh} is
\begin{IEEEeqnarray}{rCl}
  \delta & = & \delta(\amp) \eqdef
  \sigma\log\left(1+\frac{\amp}{\sigma}\right),
  \label{eq:case1choicedelta}
  \\
  \mu & = & \mu(\amp,\alpha) \eqdef \mu^{\ast} \left( 1-
    e^{-\alpha\frac{\delta^2}{2\sigma^2}} \right),
  \label{eq:case1choicemu}
\end{IEEEeqnarray}
where $\mu^{\ast}$ is the solution to \eqref{eq:3defmugurk}. For this
choice and for $\alpha=0.1$ and $0.4$ the bounds of
Theorem~\ref{thm:strongresults} are depicted in
Figures~\ref{fig:strongbounds2} and \ref{fig:strongbounds3}.

\begin{theorem}
  \label{lem:case1cor}
  If $\alpha$ lies in $\left(0,\frac{1}{2}\right)$, then  
  \begin{multline}
    \label{eq:limstrong}
    \chi(\alpha) \eqdef \lim_{\amp\uparrow\infty} \left\{
      \const{C}(\amp,\alpha\amp) - 
    \log\frac{\amp}{\sigma}\right\} \\= - \frac{1}{2}\log 2
    \pi e - (1-\alpha)\mu^{\ast} - \log(1-\alpha\mu^{\ast})
  \end{multline}
  and
  \begin{equation}\label{eq:peak_avg_lowsnr}
    \lim_{\amp \downarrow 0}\frac{\const{C}(\const{A},\alpha
      \const{A})}{\amp^2/\sigma^2}=\frac{\alpha
      (1-\alpha)}{2 }.
  \end{equation}
\end{theorem}


\subsection{Bounds on Channel Capacity with a Strong Peak-Power and
  Inactive Average-Power Constraint ($\frac{1}{2} \le \alpha \le 1$)}
\label{sec:weakresults}

By Lemma~\ref{lem:inactiveaveragepower} we have that for $\frac{1}{2}
< \alpha \leq 1$ the average-power constraint is inactive and
$\C(\amp,\alpha \amp)=\C(\amp,\frac{1}{2} \amp)$. Thus we can obtain
the results in this section by simply deriving bounds for the case
$\alpha=\frac{1}{2}$. 
\begin{theorem}
  \label{thm:weakresults}
  If $\alpha\in \left[\frac{1}{2}, 1\right]$, then
  $\const{C}(\amp,\alpha \amp)$ is lower-bounded by
  \begin{IEEEeqnarray}{rCl}
    \const{C}(\amp, \alpha \amp) & \ge & \frac{1}{2} \log \left(1+
      \frac{\amp^2}{ 2\pi e \sigma^2} \right),
    \label{eq:case2lower}
  \end{IEEEeqnarray}
  and is upper-bounded by each of the two bounds
  \begin{IEEEeqnarray}{rCl}
    \const{C}(\amp, \alpha \amp) & \le &
    \frac{1}{2}\log\left(1+\frac{\amp^2}{4\sigma^2}\right),
    \label{eq:case2upperlow}  
    \\
    \const{C}(\amp,\alpha \amp) & \le &
    \left(1-2\Qf{\frac{\delta+\frac{\amp}{2}}{\sigma}} \right)  \log 
    \frac{\amp+2\delta}
    {\sigma\sqrt{2\pi}\left(1-2\Qf{\frac{\delta}{\sigma}}\right)}
    \nonumber \\
    && -\:\frac{1}{2} + \Qf{\frac{\delta}{\sigma}}  +
    \frac{\delta}{\sqrt{2\pi}\sigma} e^{-\frac{\delta^2}{2\sigma^2}},
    \label{eq:case2upperhigh} 
  \end{IEEEeqnarray}
  where $\delta> 0$ is a free parameter. 
\end{theorem}  

\begin{figure}[tb]
  \centering
  \psfrag{A [dB]}[tc][bc]{$\frac{\amp}{\sigma}$ [dB]}
  \psfrag{C [nats]}[bc][tc]{$\const{C}$ [nats per channel use]}
  \psfrag{Weak Average-Power and Strong Peak-Power Constraint}[cb][cb]{}
  \psfrag{lower bound                                                       .}[lc][lc]{\tiny lower bound \eqref{eq:case2lower}}
  \psfrag{first upper bound}[lc][lc]{\tiny upper bound \eqref{eq:case2upperlow}}
  \psfrag{second upper bound}[lc][lc]{\tiny upper bound \eqref{eq:case2upperhigh}, using \eqref{eq:case2choicedelta}}
  \includegraphics[width=0.44\textwidth]{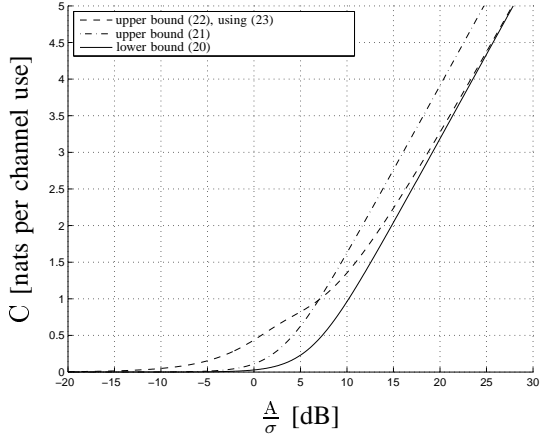}
  \caption{Bounds on the capacity of the free-space optical intensity
    channel with average- and peak-power constraints for
    $\alpha\ge\frac{1}{2}$ according to Theorem~\ref{thm:weakresults}.
    This includes the case of only a peak-power constraint $\alpha=1$.
    The free parameter has been chosen as suggested in
    \eqref{eq:case2choicedelta}. The maximum gap between upper and
    lower bound is 0.54 nats (for $\amp/\sigma\approx 7 \text{ dB}$).}
  \label{fig:weakbounds} 
\end{figure}

We suboptimally choose
\begin{equation}
  \label{eq:case2choicedelta}
  \delta = \delta(\amp) \eqdef
  \sigma\log\left(1+\frac{\amp}{\sigma}\right). 
\end{equation}
For this choice the bounds of Theorem~\ref{thm:weakresults} are
depicted in Figure~\ref{fig:weakbounds}.

\begin{theorem}
  \label{lem:case2cor}
  If $\alpha$ lies in $\left[\frac{1}{2}, 1\right]$, then
  \begin{equation}
    \label{eq:limweak}
    \chi(\alpha)\eqdef \lim_{\amp\uparrow\infty}\left\{\const{C}(\amp,
      \alpha\amp) - 
      \log\frac{\amp}{\sigma} 
    \right\} = -\frac{1}{2} \log 2\pi e
  \end{equation}
  and 
  \begin{equation}
    \label{eq:limweaklow}
    \lim_{\amp \downarrow 0 }\frac{\const{C}(\const{A},\alpha
      \amp)}{\amp^2/\sigma^2} = \frac{1}{8}.
  \end{equation}
\end{theorem}


Note that \eqref{eq:limweak} and \eqref{eq:limweaklow} exhibit the
well-known asymptotic behavior of the capacity of a Gaussian channel
under a peak-power constraint only \cite{shannon48_1}.

\subsection{Bounds on Channel Capacity with an Average-Power Constraint}
\label{sec:averageresults}

Finally, we consider the case with an average-power constraint only.

\begin{figure}[tb]
  \centering
  \psfrag{E [dB]}[tc][bc]{$\frac{\Es}{\sigma}$ [dB]}
  \psfrag{C [nats]}[bc][tc]{$\const{C}$ [nats per channel use]}
  \psfrag{Average-Power Constraint}[cb][cb]{}
  \psfrag{lower bound                                                       .}[lc][lc]{\tiny lower bound \eqref{eq:case3lower}}
  \psfrag{first upper bound}[lc][lc]{\tiny upper bound \eqref{eq:case3upperlow}, using \eqref{eq:case3choicedeltaone}, \eqref{eq:case3choicebetaone}}
  \psfrag{second upper bound}[lc][lc]{\tiny upper bound \eqref{eq:case3upperhigh}, using \eqref{eq:case3choicedeltatwo}, \eqref{eq:case3choicebetatwo}}
 \includegraphics[width=0.44\textwidth]{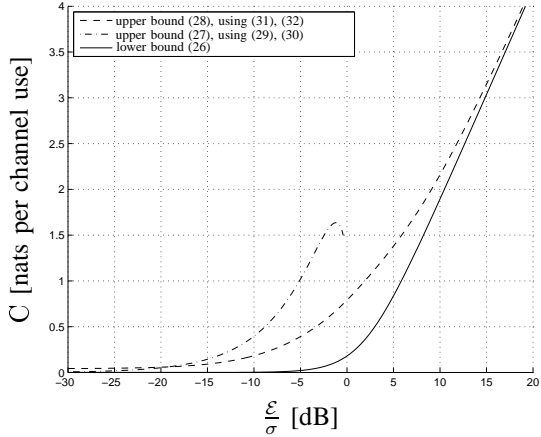}
 \caption{Bounds on the capacity of the free-space optical intensity
   channel with only an average-power constraint according to
   Theorem~\ref{thm:case3}. The free parameters have been chosen as
   suggested in
   \eqref{eq:case3choicedeltaone}--\eqref{eq:case3choicebetatwo}. The
   maximum gap between upper and lower bound is 0.64 nats (for
   $\Es/\sigma\approx 1.8 \text{ dB}$).}
  \label{fig:Avg_bounds} 
\end{figure}

\begin{theorem}
  \label{thm:case3}
In the absence of a peak-power constraint the channel capacity $\const{C}(\Es)$ is lower-bounded
  by
  \begin{IEEEeqnarray}{rCl}
    \const{C}(\Es) & \ge & \frac{1}{2} \log \left(1+ \frac{\Es^2 e}{
        2\pi \sigma^2} \right),
    \label{eq:case3lower}
  \end{IEEEeqnarray}
  and is upper-bounded by each of the bounds 
  \begin{IEEEeqnarray}{rCl}
    \const{C}(\Es) & \le &
    \log \left( \beta e^{-\frac{\delta^2}{2\sigma^2}} + \sqrt{2\pi}
      \sigma \Qf{\frac{\delta}{\sigma}}\right) - \log \left(
      \sqrt{2\pi}\sigma \right) \nonumber \\
    && -\:\frac{\delta\Es}{2\sigma^2}  +
    \frac{\delta^2}{2\sigma^2}\left( 1 - 
      \Qf{\frac{\delta}{\sigma}} -
      \frac{\Es}{\delta}\Qf{\frac{\delta}{\sigma}} \right) \nonumber
    \\
    && +\:
    \frac{1}{\beta}\left( \Es+ \frac{\sigma}{\sqrt{2\pi}} \right),
    \qquad \delta \le -\frac{\sigma}{\sqrt{e}},
    \IEEEeqnarraynumspace
    \label{eq:case3upperlow} 
    \\
    \const{C}(\Es) & \le &
    \log \left( \beta e^{-\frac{\delta^2}{2\sigma^2}} + \sqrt{2\pi}
      \sigma \Qf{\frac{\delta}{\sigma}}\right) +
    \frac{1}{2}\Qf{\frac{\delta}{\sigma}} \nonumber \\
    && +\:
    \frac{\delta}{2\sqrt{2\pi}\sigma} e^{-\frac{\delta^2}{2\sigma^2}}
    + \frac{\delta^2}{2\sigma^2}\left(1 -
      \Qf{\frac{\delta+\Es}{\sigma}}\right) \nonumber \\
    && +\:
    \frac{1}{\beta}\left(\delta+\Es+\frac{\sigma}{\sqrt{2\pi}}
      e^{-\frac{\delta^2}{2\sigma^2}} \right)  - \frac{1}{2}\log 2\pi
    e \sigma^2, \nonumber \\
    && \hfill \delta \ge 0, \quad
    \label{eq:case3upperhigh}
  \end{IEEEeqnarray}
  where $\beta>0$ and $\delta$ are free parameters. Note that bound
  \eqref{eq:case3upperlow} only holds for $\delta \le -\sigma
  e^{-\frac{1}{2}}$, while bound \eqref{eq:case3upperhigh} only holds
  for $\delta \ge 0$.
\end{theorem}

A suboptimal but useful choice for the free parameters in bound
\eqref{eq:case3upperlow} is 
shown in \eqref{eq:case3choicedeltaone} and
\eqref{eq:case3choicebetaone} and for the free parameters in bound
\eqref{eq:case3upperhigh} is 
shown in \eqref{eq:case3choicedeltatwo} and
\eqref{eq:case3choicebetatwo} at the top of the next page.
\addtocounter{equation}{4}
\begin{figure*}[!t]
  \normalsize
  \setcounter{mytempequationscounter}{\value{equation}}
  \setcounter{equation}{28}
  \begin{IEEEeqnarray}{rClCl}
    \delta & = & \delta(\Es) & \eqdef & -2\sigma\sqrt{ \log
        \frac{\sigma}{\Es}}, 
    \qquad \text{for } \frac{\Es}{\sigma} \le
  e^{-\frac{1}{4e}} \approx -0.4 \text{ dB},
    \label{eq:case3choicedeltaone}
    \\
    \beta   & = & \beta(\Es) & \eqdef & \frac{1}{2}\left( \Es +
      \frac{\sigma}{\sqrt{2\pi}}\right)
    + \frac{1}{2}\sqrt{\left( \Es +
        \frac{\sigma}{\sqrt{2\pi}}\right)^2 + 4\left(\Es +
        \frac{\sigma}{\sqrt{2\pi}}\right)\sqrt{2\pi}\sigma
      e^{\frac{\delta^2}{2\sigma^2}} \Qf{\frac{\delta}{\sigma}} },
    \label{eq:case3choicebetaone}
    \\
    \delta & = & \delta(\Es) & \eqdef & 
    \sigma \log\left(1+ \frac{\Es}{\sigma}\right),
    \label{eq:case3choicedeltatwo}
    \\
    \beta   & = & \beta(\Es) & \eqdef & \frac{1}{2}\left( \delta + \Es +
      \frac{\sigma}{\sqrt{2\pi}} e^{-\frac{\delta^2}{2\sigma^2}}\right)
    + \frac{1}{2}\sqrt{\left( \delta + \Es +
        \frac{\sigma}{\sqrt{2\pi}}
        e^{-\frac{\delta^2}{2\sigma^2}}\right)^2 + 4\left( \delta + \Es +
        \frac{\sigma}{\sqrt{2\pi}}
        e^{-\frac{\delta^2}{2\sigma^2}}\right)\sqrt{2\pi}\sigma
      e^{\frac{\delta^2}{2\sigma^2}} \Qf{\frac{\delta}{\sigma}} }.
    \IEEEeqnarraynumspace
    \label{eq:case3choicebetatwo}
  \end{IEEEeqnarray}
  \setcounter{equation}{\value{mytempequationscounter}}
  \hrulefill
  \vspace*{4pt}
\end{figure*}
For these choices, the bounds of Theorem~\ref{thm:case3} are depicted
in Figure~\ref{fig:Avg_bounds}.

\begin{theorem}
  \label{lem:case3cor}
  In the case of only an average-power constraint,  
  \begin{equation}\label{eq:limEs}
    \chi_{\Es}\eqdef \lim_{\Es\uparrow\infty} \left\{\const{C}(\Es) -
      \log \frac{\Es}{\sigma} \right\} =  \frac{1}{2}\log
    \frac{e}{2\pi} 
  \end{equation}
  and
  \begin{IEEEeqnarray}{rCl}
    \varlimsup_{\Es \downarrow 0}
    \frac{\const{C}(\Es)}{\frac{\Es}{\sigma}  
      \sqrt{ \log\frac{\sigma}{\Es}}} & \leq  & 2,
    \\
    \varliminf_{\Es \downarrow 0}
    \frac{\const{C}(\Es)}{\frac{\Es}{\sigma} 
      \sqrt{ \log\frac{\sigma}{\Es}}} & \geq & \frac{1}{ \sqrt{2}}.  
  \end{IEEEeqnarray}
\end{theorem}
Note that the asymptotic upper and lower bound at low SNR do not
coincide in the sense that their ratio equals $2 \sqrt{2}$ instead of
1. However, they exhibit similar behavior. 


\addtolength{\textheight}{-0.2cm}

\section{Derivation}
\label{sec:derivation}

In the following we will outline the derivations of the firm lower and
upper bounds given in the previous section.

One easily finds a lower bound on capacity by dropping the
maximization and choosing an arbitrary input distribution $Q(\cdot)$
to compute the mutual information between input and output.  However,
in order to get a tight bound, this choice of $Q(\cdot)$ should yield
a mutual information that is reasonably close to capacity. Such a
choice is difficult to find and might make the evaluation of $I(X;Y)$
intractable. The reason for this is that even for relatively ``easy''
distributions $Q(\cdot)$, the corresponding distribution on the
channel output $Y$ may be difficult to compute, let alone $h(Y)$.  We
circumvent these problems by using the entropy power inequality
\cite[Th.~17.7.3]{coverthomas06_1} to lower-bound $h(Y)$ by an
expression that depends only on $h(X)$. \emph{I.e.}, we ``transfer''
the problem of computing (or bounding) $h(Y)$ to the input side of the
channel, where it is much easier to choose an appropriate distribution
that leads to a tight lower bound on channel capacity:
\begin{IEEEeqnarray}{rCl}
  \const{C} & = & \sup_{Q(\cdot)} I(X;Y)
  \\
  & \ge & I(X;Y) \big|_{\text{for a specific $Q(\cdot)$}}
  \\
  & = & \big(h(Y) - h(Y|X) \big)\big|_{\text{for a specific $Q(\cdot)$}}
  \\
  & = & h(X+Z)\big|_{\text{for a specific $Q(\cdot)$}} - h(Z) 
  \\
  & \ge & \left. \frac{1}{2}\log\left(e^{2h(X)}+e^{2h(Z)}\right)
  \right|_{\text{for a specific $Q(\cdot)$}} - h(Z)
  \label{eq:pei}
  \IEEEeqnarraynumspace
  \\
  & = & \left. \frac{1}{2}\log\left(1+ \frac{e^{2h(X)}}{2\pi e
        \sigma^2}\right) \right|_{\text{for a specific $Q(\cdot)$}}
  \label{eq:lowerbound}
\end{IEEEeqnarray}
where the inequality in \eqref{eq:pei} follows from the entropy power
inequality.  To make this lower bound as tight as possible we will
choose a distribution $Q(\cdot)$ that maximizes differential entropy
under the given constraints \cite[Ch.~12]{coverthomas06_1}.

The derivation of the upper bounds in Section~\ref{sec:3resultsone}
are based on the duality approach \eqref{eq:upperbound1}.  Hence, we
need to specify a distribution $R(\cdot)$ and evaluate the relative
entropy in \eqref{eq:upperbound1}.

We have chosen output distributions $R(\cdot)$ with the following
densities. For \eqref{eq:case1upperlow} we choose
\begin{equation}
  R'(y)\eqdef \frac{1}{\sqrt{2\pi\left(\sigma^2+\Es(\amp-\Es)\right)}}
  e^{-\frac{\left(y-\Es\right)^2}{2\sigma^2+2\Es(\amp-\Es)}};
\end{equation}
for \eqref{eq:case1upperhigh} we choose
\begin{equation} 
  \label{eq:densitycase1high}
  R'(y)\eqdef  \begin{cases}
    \frac{1}{\sqrt{2\pi}\sigma}e^{-\frac{y^2}{2\sigma^2}}, & y <
    -\delta, 
    \\[3mm]
    \frac{1}{\amp}\cdot\frac{\mu \left(1 -
        2\Qf{\frac{\delta}{\sigma}}\right) 
    }{e^{\frac{\mu\delta}{\amp}}-e^{-\mu\left(1+\frac{\delta}{\amp}\right)}}
    e^{-\frac{\mu y}{\amp}}, & -\delta \leq y \leq \amp+\delta,  
    \\[3mm] 
    \frac{1}{\sqrt{2\pi}\sigma}e^{-\frac{(y-\amp)^2}{2\sigma^2}}, & y>
    \amp+\delta,  
  \end{cases}
\end{equation}
where $\delta> 0$ and $\mu>0$ are free parameters;
for \eqref{eq:case2upperlow} we choose
\begin{equation}
  \label{eq:lowPeak_R}
  R'(y)\eqdef \frac{1}{\sqrt{2\pi\left(\sigma^2+\frac{\amp^2}{4}\right)}}
  e^{-\frac{\left(y-\frac{\amp}{2}\right)^2}{2\sigma^2+\frac{\amp^2}{2}}};
\end{equation}
for \eqref{eq:case2upperhigh} we choose
\begin{equation} 
  \label{eq:HighSNRPeak_R}
  R'(y)\eqdef  \begin{cases}
    \frac{1}{\sqrt{2\pi}\sigma}e^{-\frac{y^2}{2\sigma^2}}, & y <
    -\delta, 
    \\[2mm]
    \frac{1-2\Qf{\frac{\delta}{\sigma}}}{\amp+2 \delta}, & -\delta \leq y \leq
    \amp+\delta,  
    \\[2mm]  
    \frac{1}{\sqrt{2\pi}\sigma}e^{-\frac{(y-\amp)^2}{2\sigma^2}}, & y>
    \amp+\delta,  
  \end{cases}
\end{equation}
where $\delta> 0$ is a free parameter;
and  for \eqref{eq:case3upperlow} and \eqref{eq:case3upperhigh}
we choose
\begin{equation} 
  \label{eq:densitycase3low}
  R'(y)\eqdef  \begin{cases}
    \frac{1}{\beta e^{-\frac{\delta^2}{2\sigma^2}} +
      \sqrt{2\pi}\sigma\Qf{\frac{\delta}{\sigma}}} e^{-\frac{y^2}{2\sigma^2}}, & y <
    -\delta, 
    \\[3mm]
    \frac{1}{\beta e^{-\frac{\delta^2}{2\sigma^2}} +
      \sqrt{2\pi}\sigma\Qf{\frac{\delta}{\sigma}}}
    e^{-\frac{\delta^2}{2\sigma^2}}
    e^{-\frac{y+\delta}{\beta}}, & y \ge -\delta,
  \end{cases}
\end{equation}
where $\delta \in \Reals$ and $\beta>0$ are free parameters. In the
derivation of \eqref{eq:case3upperhigh} we then restrict $\delta$ to
be nonnegative, while in the derivation of \eqref{eq:case3upperlow} we
restrict $\delta \le -\sigma e^{-\frac{1}{2}}$.

\section*{Acknowledgments}

The authors would like to thank Ding-Jie Lin for fruitful discussions.
The work of S.~M.~Moser was supported in part by the ETH under TH-23
02-2.



\end{document}